\documentclass[modern]{aastex63}

\def\gtsima{$\;\buildrel > \over \sim \;$}
\def\simgt{\lower.5ex \hbox{\gtsima}}
\def\ltsima{$\;\buildrel < \over \sim \;$}
\def\simlt{\lower.5ex \hbox{\ltsima}}

\begin{document}

\title{H$_2$ fluorescent emission from the diffuse interstellar medium}

\author{David A. Neufeld}
\affiliation{William H.\ Miller Department of Physics \& Astronomy, Johns Hopkins University, Baltimore, MD 21218, USA}

\begin{abstract}

A simple analytic model is presented to predict the near-IR H$_2$
fluorescent line intensities emitted by diffuse interstellar clouds
with a Plummer density profile.  It is
applicable to sightlines where (1) the column densities of H and H$_2$ have
been measured and (2) the peak gas density can be estimated from extinction maps or
observations of C$_2$ absorption.

\end{abstract}

\section{Introduction}

Several recent developments motivate renewed interest in the H$_2$ fluorescent
emissions expected from the diffuse interstellar medium.  
Three-dimensional 
extinction maps (Edenhofer et al.\ 2024, and references therein)
based on recent {\it Gaia} observations 
provide unique information about
the density structure of the interstellar medium.
They suggest that cold diffuse clouds, rather than being at
constant density, show a centrally-peaked density profile that is well fit
(Zucker et al.\ 2021; hereafter Z21) by a double Gaussian or by a
Plummer profile of
the form $n_{\rm H}(r) = n_0 (1 + r^2/r_0^2)^{-p/2}$,
where $n_{\rm H}=n({\rm H}) + 2 n({\rm H_2})$ 
is the density of H nuclei, $n_0$ is the density at the cloud center, $r$ is
the distance from the cloud center, and $r_0$ is a constant.
In a sample of nearby molecular clouds, Z21 obtained a median power-law index, $p$, of
1.8 (with a range of 1.2 -- 3.4) and a median peak density, $n_c$, of 49~cm$^{-3}$ 
(with a range of 35 to 117~cm$^{-3})$.  For a ray passing though such a cloud with 
impact parameter, $b$, 
the one-dimensional density profile along the sightline also has a Plummer profile:
$$n_{\rm H}(z) = {n_{\rm max} \over (1 + z^2/z^2_0)^{p/2}}, \eqno(1)$$
with $z_0 = (r_0^2+b^2)^{1/2}$ and $n_{\rm max} = n_0 (1 + b^2/r_0^2)^{-p/2}$. 
Here $n_{\rm max}$ is the peak density along the sightline and $z$ is the distance 
along the sightline from the location where that peak occurs.

Along the sightlines to hot stars, ultraviolet
absorption-line observations can be used to measure the column densities of
atomic and molecular hydrogen, $N({\rm H})$ and $N({\rm H}_2)$, providing 
valuable ancillary information about the moelcular fraction
(e.g.\ Obolentseva et al. 2024, and references therein).
Moreover, optical and UV observations of individual 
C$_2$ rotational states -- interpreted using newly-available molecular data
(Najar \& Kalugina 2020) -- can provide an independent estimate of the peak density of
H nuclei, $n_{\rm max}$, along such sightlines;
for nearby clouds, where the extinction maps are most reliable, such C$_2$-derived
density estimates are in good agreement with those offered by the extinction maps
(Neufeld et al.\ 2024).

Near-IR spectroscopy, which can now be performed at unprecedented sensitivity with the
NIRSpec instrument on JWST, will offer another potential probe of diffuse clouds through
the observation of H$_2$ fluorescent emissions.  Whereas previous observations of 
interstellar H$_2$ fluorescent emissions were primarily 
confined to dense clouds lying close to 
a source of UV radiation (e.g. Le et al 2017), JWST opens the possibility of detecting
such emissions from diffuse molecular clouds exposed to
the average radiation field in the Galaxy.

In this research note, I present a simple analytic model that may be used
to obtain an estimate of the fluorescent H$_2$ emissions along sightlines
of known $N({\rm H})$, $N({\rm H}_2)$, and $n_{\rm max}$.

\section{Calculation}

For cold gas at densities below $\sim 10^3\,\rm cm^{-3}$, the intensity of
H$_2$ fluorescent emission is a measure of the photodissociation rate,
both processes being caused by the absorption of ultraviolet photons in the 
Lyman and Werner bands of H$_2$. 
For gas in chemical equilibrium, with the $\rm H_2$ destruction rate dominated
by photodissociation and balanced by the formation rate via grain catalysis,
the fluorescent intensity, in any given $\rm H_2$ line is given by
$$I_f = {(1 - f_a) \over 4 \pi} \int R n_{\rm H} n({\rm H}) E dz, \eqno(2)$$
(Neufeld \& Spaans 1996, hereafter NS96) 
where $z$ is the distance along the line-of-sight, $R$ is the rate coefficient for H$_2$ formation 
on grains, $E$ is the
mean energy per dissociation emitted in the line under consideration, 
and $f_a$ is the fraction of the emitted radiation that is absorbed by dust
before it reaches us. 
If we approximate
$R$ and $E$ as constant along the sightline (NS96), we obtain
$I_f =  (1 - f_a) RN({\rm H}) \bar{n} E / (4 \pi)$, where 
$\bar{n}$ is the $n({\rm H})$-weighted average of $n_{\rm H}$, 
$$\bar n = {1 \over N({\rm H})}{\int n_{\rm H} n({\rm H}) dz}. \eqno(3)$$

For a uniform cloud of constant density, $\bar{n}$
is trivially equal to the peak density, $n_{\rm max}$, but for a centrally-peaked
density distribution $\bar{n}$ is invariably less than $n_{\rm max}$.
For a Plummer density profile, the solution is relatively simple 
for the specific case where $p=2$.
The total column density of H nuclei, $N_{\rm H}$, is given by
$$N_H =  \int {n_{\rm max} dz \over 1 + z^2/z^2_0} = \pi n_{\rm max} z_0. \eqno(4)$$ 
Because of the increasing density and shielding H$_2$ column, 
the molecular fraction increases rapidly as $\vert z \vert$ decreases.
This motivates a Stromgren-like approximation (SLA), in which the gas 
is assumed to be fully molecular for $\vert z \vert < z_s$ and 
fully atomic for $\vert z \vert > z_s$.  With this approximation,
the $\rm H_2$ column density is 
$$N({\rm H}_2) =  \onehalf  \int_{-z_s}^{+z_s} {n_{\rm max} dz \over (1 + z^2/z^2_0)} = 
n_{\rm max} z_0 \,{\rm tan}^{-1}\,(z_s/z_0). \eqno(5)$$
Thus $z_s$ is related to the average
molecular fraction along the sightline, $f_{\rm H_2} = 2N({\rm H}_2)/N_{\rm H},$
by the expression $$f_{\rm H_2} = {2\, {\rm tan}^{-1}(z_s/z_0) \over \pi},\eqno(6)$$
or equivalently $z_s = z_0 {\rm tan}(\pi f_{\rm H_2}/2)$.
We may now obtain $\bar{n}$ in the form
$$\bar{n} = {1 \over N({\rm H})} \int n_{\rm H}^2 x({\rm H}) dz = 
{n_{\rm max}  \over N({\rm H})} \int {x({\rm H}) n_{\rm max} dz \over (1 + z^2/z^2_0)^2}\eqno(7)$$ 
where $x({\rm H)}=n({\rm H})/n_{\rm H}.$  
With the SLA, we obtain
$${\bar{n} \over n_{\rm max}} = {2 \over N({\rm H})} \int_y^\infty {n_{\rm max} dz \over (1 + z^2/z^2_0)^2} 
=\biggl[ {\pi \over 2} - {y \over 1+y^2} - {\rm tan}^{-1}y \biggr] n_{\rm max} z_0,\eqno(8)$$
where $y = z_s/z_0 = {\rm tan}(\pi f_{\rm H_2}/2).$
Substituting for this expression for $y$ into equation (8), and making use of
well-known trigonometric identities, we finally obtain
$${\bar{n} \over n_{\rm max}} = 
\onehalf - \biggl({{\rm sin}(\pi f_{\rm H_2}) \over 2 \pi (1 - f_{\rm H_2})} \biggr) = {1 - {\rm sinc}(\pi[1-f_{\rm H_2}])  \over 2}\eqno(9)$$

For Plummer indices, $p$, other than 2, the computation of $\bar{n}$ is more 
complicated, but analytic solutions to the relevant integrals
can still be found in terms of the Gaussian (a.k.a.\ ordinary) hypergeometric function
$_2F_1(a,b;c;d)$ (SciPy special function {\tt scipy.special.hyp2f1}).  The generalizations 
of equations (6) and (8) are

%2.*y*sp.hyp2f1(0.5,0.5*p,1.5,-y*y)*sp.gamma(0.5*p)/(np.sqrt(np.pi)*sp.gamma(0.5*p-0.5))
%	xnumer = sp.gamma(    p-0.5)*np.sqrt(np.pi)/(2.*sp.gamma(    p))-y*sp.hyp2f1(0.5,    p,1.5,-y*y)
%	xdenom = sp.gamma(0.5*p-0.5)*np.sqrt(np.pi)/(2.*sp.gamma(0.5*p))-y*sp.hyp2f1(0.5,0.5*p,1.5,-y*y)
$$f_{\rm H_2} = 
 {2 \over \sqrt \pi} \biggl({_2F_1 (\onehalf,{p \over 2};{3 \over 2};-y^2) \,y\Gamma({p \over 2}) \over  \Gamma({p \over 2}-\onehalf)}\biggr), \eqno(10)$$
and 
$${\bar{n} \over n_{\rm max}} = {_2F_1 (\onehalf,p;{3 \over 2};-y^2)\,y - \onehalf \Gamma(p-\onehalf) \pi^{1/2}/\Gamma(p)     \over
                         _2F_1 (\onehalf,{p \over 2};{3 \over 2};-y^2)\,y - \onehalf \Gamma({p \over 2}-\onehalf) \pi^{1/2}/\Gamma({p \over 2}) }. \eqno(11)$$

\section{Results}

Figure 1 (upper panel) shows the quantity $\bar{n}/n_{\rm max}$ (eqn.\ 11) as a function of
the sightline-averaged molecular fraction, $f_{\rm H_2}$ (eqn.\ 10).  Results are
shown for power-law indices of 1.6, 2.0, 2.4, 2.8, and 3.2 (magenta, cyan, blue, green, 
and red curves respectively).  The results given for $p=2$ by equation (9) are overplotted
with the black dashed curve as a check.  The value of $\bar{n}/n_{\rm max}$ is a decreasing
function of $f_{\rm H_2}$, a behavior that reflects the fact that the atomic region moves
outward to gas of lower density as the molecular fraction increases.  The use of the 
SLA to obtain these results means that they may be regarded as lower limits on 
the actual $\bar{n}/n_{\rm max}$: if
$x({\rm H})$ exceeds zero for $\vert z \vert < z_s$, then more of the
atomic hydrogen column is located at higher densities than was assumed.

\begin{figure}[t!]
\includegraphics[angle=0,width=6.3 true in]{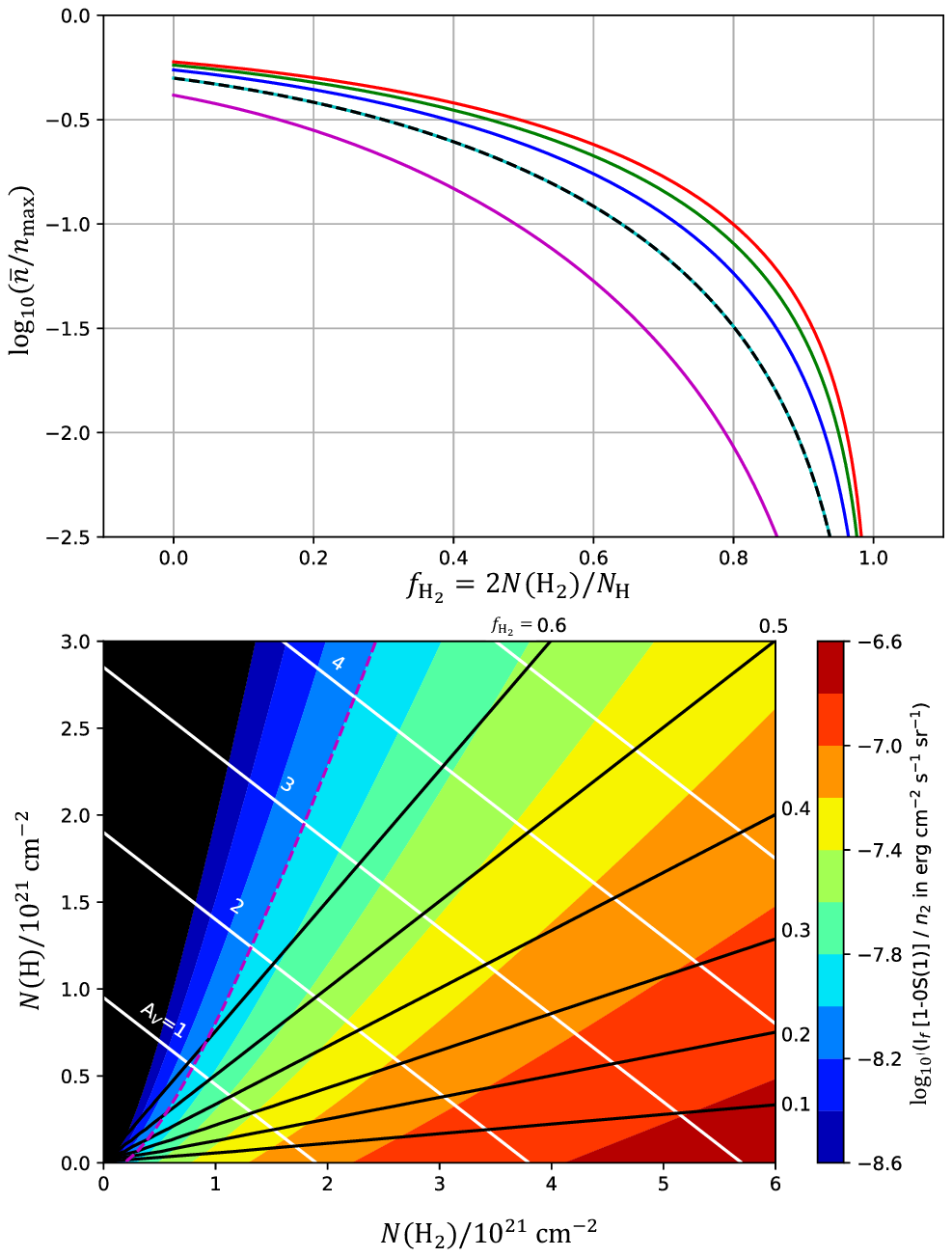}
\figcaption{Results of the analytic treatment presented here (see the text for
details).  Upper panel: $\bar{n}/n_{\rm max}$ versus $f_{\rm H_2}$, for Plummer density profiles 
with power-law indices, $p$, 
of 1.6, 2.0, 2.4, 2.8, and 3.2 (magenta, cyan, blue, green, 
and red curves respectively).  Lower panel: predicted fluorescent line intensities for 
the $v=1-0 S(1)$ transition at 2.122$\,\mu$m, divided by $n_2$.  }
\end{figure}

The lower panel of Figure 1 shows predictions for 
$${I_f \over n_2} =  10^2\,{\rm cm}^{-3} (1 - f_a) {RN({\rm H}) \, E \over 4 \pi}\biggl({\bar{n} \over n_{\rm max}}\biggr),\eqno(12)$$ 
where $n_2 =n_{\rm max}/[10^2\,\rm cm^{-3}].$  
They apply to the v=1-0 S(1) transition, and are shown in the 
plane of the two key observables $N({\rm H})$ and $N({\rm H}_2)$.
To obtain the results shown in this plot, 
I adopted the expression for $(1 - f_a)$ given by NS96; 
a value for $R$ of $3 \times 10^{-17}\rm \,cm^3\,s^{-1}$; 
and a value for $E$ of $4.6 \times 10^{-13}\,\rm erg$, the
latter having been computed by NS96 for a typical H$_2$ ortho-to-para ratio of unity.
Parallel white lines show the loci of constant line-of-sight extinction,
$A_V$, for an assumed $N_{\rm H}/A_V$ ratio of $1.9 \times 10^{21}\, \rm cm^{-2}$ per mag,
with the $A_V$ values denoted by white numerals.  Loci of constant $f_{\rm H_2}$ are also 
shown in black with their $f_{\rm H_2}$ values labeled with red numerals.  For $n_2=1$, 
the dashed magenta line
shows the typical JWST/NIRSpec detection limit with the 
multishutter array (Padovani et al.\ 2022) of  
$10^{-8} \, \rm erg\,cm^{-2}\,s^{-1}\,sr^{-1}$ (3 $\sigma$ in 1.25 hr using 25 shutters).

The results presented here apply specifically to gas in chemical equilibrium.
As noted by Goldshmidt \& Sternberg (1995) and by Bialy et al.\ (2024), 
if the H$_2$ abundance is out of equilibrium and increasing (decreasing) 
with a destruction rate smaller (larger) 
than the formation rate, then the expected line intensity will be smaller (larger) than
the values plotted in the lower panel of Figure 1.

I am pleased to acknowledge helpful discussions with S.\ Bialy.

\end{document}